\documentclass[aps,floatfix,twocolumn,final,prl]{revtex4-2}
\usepackage{epsfig}
\usepackage{graphicx}
\usepackage[ansinew]{inputenc}
\usepackage{array}
\usepackage{hyperref}
\usepackage{tabularx}

\usepackage{xcolor}
\usepackage{ulem}

\newcolumntype{C}[1]{>{\centering\let\newline\\\arraybackslash\hspace{0pt}}m{#1}}

\begin{document}

\preprint{APS/123-QED}
\title{Microscopic probe of magnetic polarons in antiferromagnetic Eu$_{5}$In$_{2}$Sb$_{6}$}

\author{J. C. Souza$^{1}$, S. M. Thomas$^{2}$, E. D. Bauer$^{2}$, J. D. Thompson$^{2}$, F. Ronning$^{2}$, P. G. Pagliuso$^{1}$ and P. F. S. Rosa$^{2}$}

\affiliation{$^{1}$Instituto de F\'isica \lq\lq Gleb Wataghin\rq\rq,
UNICAMP, 13083-859, Campinas, SP, Brazil\\
$^{2}$Los Alamos National Laboratory, Los Alamos, New Mexico 87545, USA}


\date{\today}

\begin{abstract}
Colossal magnetoresistance (CMR) emerges from intertwined spin and charge degrees of freedom in the form of ferromagnetic clusters also known as trapped magnetic polarons. As a result, CMR is rarely observed in antiferromagnetic materials. Here we use electron spin resonance (ESR) to reveal microscopic evidence for the formation of magnetic polarons in antiferromagnetic Eu$_{5}$In$_{2}$Sb$_{6}$. First, we observe a reduction of the Eu$^{2+}$ ESR linewidth as a function of the applied magnetic field consistent with ferromagnetic clusters that are antiferromagnetically coupled. Additionally, the Eu$^{2+}$ lineshape changes markedly below $T'$ $\sim$ 200 K, a temperature scale that coincides with the onset of CMR. The combination of these two effects provide strong evidence that magnetic polarons grow in size below $T'$ and start influencing the macroscopic properties of the system.
\end{abstract}

\maketitle

\section{\label{sec:intro}I. Introduction}

Low-carrier density materials containing divalent europium are prone to exhibiting colossal magnetoresistance (CMR) \cite{kasuya1968anomalous,ramirez1997colossal}, and one explanation for this phenomenon is the formation of magnetic polarons \cite{pohlit2018evidence}. These quasiparticles are a result of free carriers at low densities that self-trap in ferromagnetic clusters around Eu$^{2+}$ local moments \cite{kaminski2002polaron}. There are several Eu-based materials with different crystal structures and magnetic ground states that host these emergent quasiparticles, but most of these systems display ferromagnetic order at low temperatures \cite{pohlit2018evidence,sullow2000magnetotransport,devlin2018eu11zn4sn2as12,salamon2001physics,oliveira1972eute,shapira1972eute}.

With the advent of nontrivial topology in condensed-matter physics, a question that naturally arises is the role of topology in strongly correlated systems such as those exhibiting CMR \cite{tokura2017emergent,paschen2020quantum}. Further, the interplay between magnetism and nontrivial band topology is predicted to give rise to emergent topological quantum phenomena, such as the axion insulating state in MnBi$_{2}$Te$_{4}$ \cite{li2019intrinsic,zhang2019topological,hao2019gapless,deng2020quantum}. Just like Mn$^{2+}$ in MnBi$_{2}$Te$_{4}$, Eu$^{2+}$ could play a similar role in a nontrivial background. Motivated by the prediction of topological phases in nonsymmorphic crystal structures \cite{parameswaran2013topological}, here we focus on nonsymmorphic Eu$_{5}$In$_{2}$Sb$_{6}$, which has been recently synthesized in single crystalline form \cite{rosa2020colossal}. Band structure calculations for the uncorrelated analog Ba$_{5}$In$_{2}$Sb$_{6}$ reveal conflicting results regarding its topology \cite{wang2016hourglass,bradlyn2017topological,wieder2018wallpaper,vergniory2019complete,zhang2019catalogue,tang2019comprehensive}, and a microscopic investigation of these materials is imperative.

Orthorhombic Eu$_{5}$In$_{2}$Sb$_{6}$ has three distinct Eu$^{2+}$ sites, which give rise to two antiferromagnetic (AFM) transitions at $T_{N1}$~=~14 K and at $T_{N2}$ = 7 K. Eu$_{5}$In$_{2}$Sb$_{6}$ also exhibits CMR that peaks at - 99.999\% for magnetic fields of $H~=~9$~T and temperatures just above $T_{N1}$. Recent studies argue that magnetic polarons start to form below $T'$ $\approx$ 210 K, a temperature scale characterized by different macroscopic signatures: a deviation from the Curie-Weiss law, the appearance of an anomalous Hall effect, and the onset of CMR \cite{rosa2020colossal}. 

The strong exchange interaction between conduction electrons and Eu$^{2+}$ 4$f$ moments combined with the lack of an orbital moment in Eu$^{2+}$ ions makes Eu$_{5}$In$_{2}$Sb$_{6}$ an ideal testbed to be explored by electron spin resonance (ESR).  The ESR linewidth $\Delta H$ is inversely proportional to the spin-spin relaxation time $T_{2}$, which can be affected by internal fields, a distribution of exchange interactions, and the spin-flip scattering between the 4$f$ moments and conduction electrons \cite{rettori1973electron,barnes1981theory,abragam2012electron}. The latter is also known as the Korringa mechanism. Additionally, the $g$-value, which is proportional to the resonance field $H_{r}$, also gives information about internal fields and their nature as well as the interaction between 4$f$ moments and the conduction electrons \cite{barnes1981theory,abragam2012electron}. In systems with magnetic polarons, the spin-flip scattering is reduced as a function of the applied magnetic field $H$, which is accompanied by a negative MR due to the increase of the quasiparticle size. Consequently, $T_{2}$ increases and there is an unusual narrowing of $\Delta H$ as a function of $H$ \cite{urbano2004magnetic}. ESR has been previously employed to investigate magnetic polarons in EuB$_{6}$ \cite{urbano2004magnetic} and Eu-based clathrates \cite{rosa2013magnetic} as well as in $d$-electron systems in combination with muon spin rotation \cite{yang2004magnetic,choi2010inhomogeneous,li2018evidence,storchak2010spin}.

In this Letter, we report a microscopic ESR investigation of Eu$_{5}$In$_{2}$Sb$_{6}$ \cite{rosa2020colossal} as a function of different microwave frequencies and crystallographic directions. The reduction of the Eu$^{2+}$ ESR linewidth as a function of $H$ is a microscopic fingerprint of the presence of ferromagnetic clusters. Importantly, below $T'$~$\approx$~200~K a marked change of the Eu$^{2+}$ ESR lineshape is driven by the increase of the microwave skin depth effect. These two experimental results not only provide evidence of the presence of magnetic polarons in this system but also demonstrate that polarons start to influence the macroscopic properties of the system at $T'$ $\approx$ 200 K.

Importantly, the Eu$^{2+}$ spin dynamics sheds light on the relevant magnetic exchange interactions in Eu$_{5}$In$_{2}$Sb$_{6}$. The Eu$^{2+}$ ESR $\Delta H$ angle dependence reveals an anisotropy in the spin-flip scattering that resembles that of the electrical resistivity, which suggests that the observed angular dependence is connected to a residual Fermi surface anisotropy. The Eu$^{2+}$ ESR temperature dependent $g$-value shows a slight but systematic reduction as a function of $H$, which corroborates the presence of AFM correlations between polarons. Finally, the temperature and angle dependencies of the $g$-value for different crystallographic directions reveals an interplay of FM and AFM short-range interactions in the $ab$ plane and a FM component along the $c$ axis.

\section{\label{sec:meth}II. Methods}

Single crystalline samples of Eu$_{5}$In$_{2}$Sb$_{6}$ were synthesized by a combined In-Sb flux described elsewhere \cite{rosa2020colossal}. Resistivity measurements were performed using a four-probe configuration in a commercial low-frequency AC bridge in voltage mode with an applied voltage of $V$ = 200 $\mu$V. ESR measurements were performed for single crystals in X- ($\nu$ = 9.4 GHz) and Q-bands ($\nu$ = 34 GHz) commercial spectrometers equipped with a goniometer and a He-flow cryostat in the temperature range of 15 K $\leq$ $T$ $\leq$ 300 K. The crystals have a rod-like shape, and the $c$ axis is the long axis. Typical sample sizes are 0.5 mm x 0.5 mm x 3 mm. Particular care was taken to avoid any possible sample size effects in the ESR measurements. Due to the smaller cavity and the crystal shape we were not able to obtain Q-band measurements for fields applied parallel to the $c$ axis. The ESR spectra were analyzed using the software Spektrolyst.

\section{\label{sec:resultsanddiscussion}III. Results and Discussion}

\begin{figure}[!ht]
\includegraphics[width=\columnwidth]{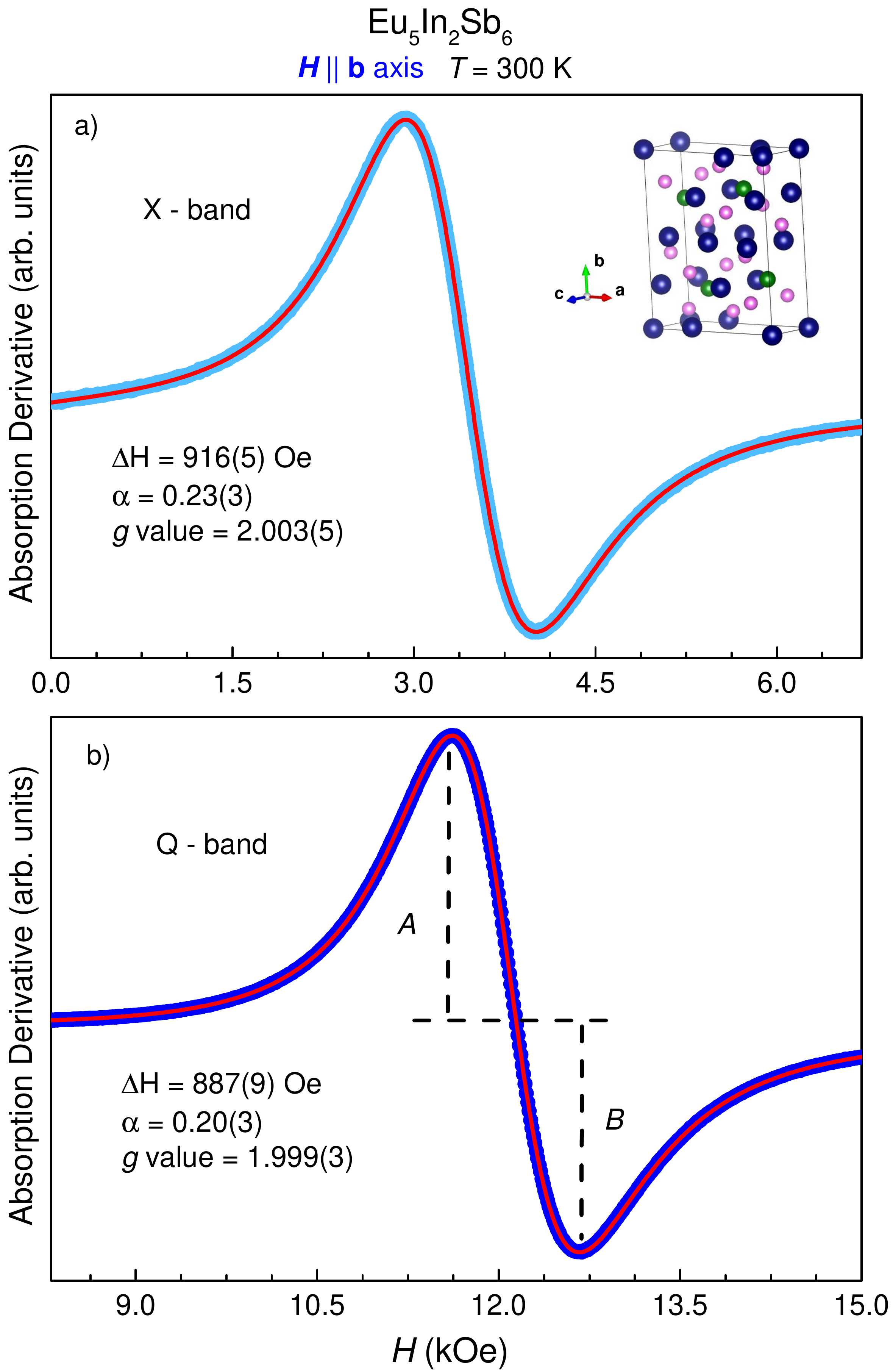}
\caption{a) X- and b) Q-bands Eu$^{2+}$ ESR spectra for Eu$_{5}$In$_{2}$Sb$_{6}$. $H$ is applied parallel to the $b$ axis. The right inset in the top panel shows the crystalline structure. The $A$($B$) showed in the lower panel denotes the distance from the base line to the peak (valley). The red solid lines are fits explained in the text.}
\label{Fig1}
\end{figure}

Figures \ref{Fig1} a) and b) show the Eu$^{2+}$ ESR spectra with $H$ parallel to the $b$ axis at $T$ = 300 K for X- and Q-bands, respectively. The $b$ axis is identified in the inset of Fig. \ref{Fig1} a). The red solid lines are the best fits to the Eu$^{2+}$ ESR spectra, represented by the power absorption derivative (d$P$/d$H$) as a function of $H$:

\begin{equation}
\frac{dP}{dH} \propto (1-\alpha)\frac{\mathrm{d} }{\mathrm{d} x}\left ( \frac{1}{1+x^2} \right 
) + \alpha \frac{\mathrm{d} }{\mathrm{d} x} \left ( \frac{x}{1+x^2} \right ),
\label{Eq1}
\end{equation}
where $\alpha$ is the asymmetric parameter of the line shape and $x$ = $2(H - H_{r})/\Delta H$ \cite{feher1955electron}. As shown in Fig. \ref{Fig1}, there is a reduction of the Eu$^{2+}$ ESR $\Delta H$ from low (X-band) to high frequency (Q-band) even at room temperature. This reduction reveals that the Eu$^{2+}$ resonance is homogeneous in the paramagnetic state, indicating the good quality of the samples, and that there is a reduction of the spin-flip scattering as a function $H$.

The asymmetry of the Eu$^{2+}$ ESR line shape, which will be reflected in the A/B ratio defined in Fig. \ref{Fig1} b), is a consequence of the microwave skin depth $\delta$ = $\sqrt{{\rho}/{\pi \nu \mu_{0} \mu{r}}}$, wherein $\rho$ is the resistivity, $\mu_{0}$ the vacuum permeability, and $\mu_{r}$ the relative permeability. The skin depth, in a semi-classical view, is a consequence of the local shielding currents of carriers driving electromagnetic fields out of phase \cite{hemmida2018weak,barnes1981theory,poole1971relaxation,abragam2012electron}. In other words, the evolution of $\delta$ is a reflection of the evolution of the local AC resistivity around our Eu$^{2+}$ probe \cite{urbano2002different,urbano2005gradual,souza2020metallic}. If the sample size $L$ is much smaller than $\delta$, one obtains a symmetric Lorentzian line shape (A/B = 1; $\alpha$ = 0). In a metallic environment, the thickness to skin depth ratio $\lambda$ = $L$/$\delta$ $\gg$ 1, and one obtains a Dysonian line shape (A/B $\approx$ 2.7, $\alpha$ $\approx$ 0.55) \cite{feher1955electron,pake1948line}.

The Dyson model for the $\lambda$-dependence of $\alpha$ considers the microwave penetration in a flat plate \cite{dyson1955electron,pifer1971conduction}. In this model, $\alpha$ is given by:

\begin{equation}
\alpha = \frac{\mathrm{sinh}^{2} \lambda - \mathrm{sin}^{2} \lambda}{(\mathrm{cosh} \lambda + \mathrm{cos} \lambda)^{2}}.
\label{Eq2}
\end{equation}

\begin{figure}[!ht]
\includegraphics[width=\columnwidth]{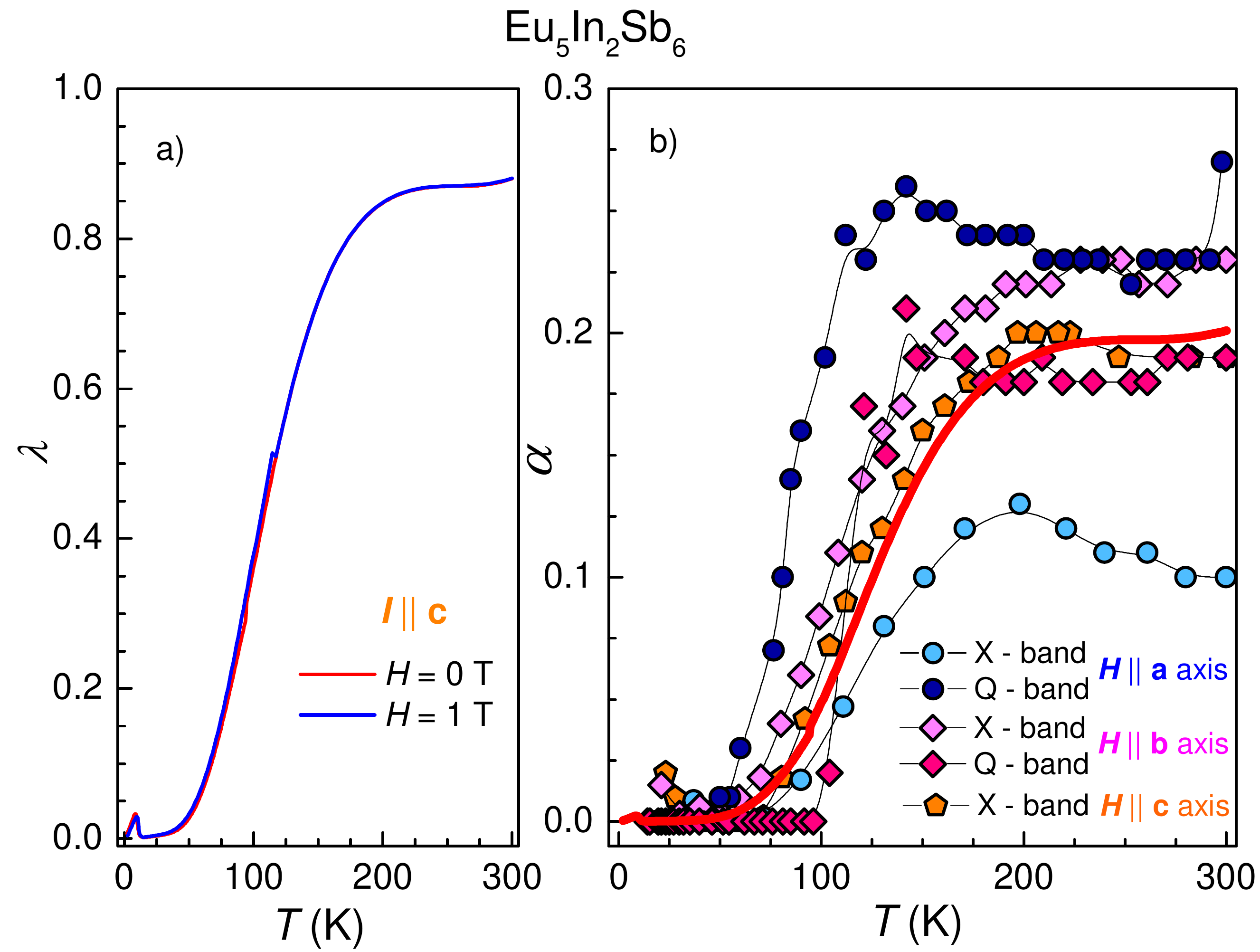}
\caption{a) $\lambda$ for current $I$ applied parallel to the $c$ direction and b) $\alpha$ for different directions and microwave frequencies as a function of temperature. The red solid line is a simulation described into the text.}
\label{Fig2}
\end{figure}

Figure \ref{Fig2} a) shows the $T$-dependence of $\lambda$, which is derived from DC electrical resistivity, in Eu$_{5}$In$_{2}$Sb$_{6}$ for two different $H$s. Because the difference between zero field and 1~T is negligible, we used the zero-field resistivity to simulate the $T$-dependence of $\alpha$ for the X-band, as shown by the red solid line in Fig. \ref{Fig2} b). Below 
$T'~\approx 200$~K, there is a clear decrease of $\alpha$ for all crystallographic axes. This temperature scale coincides with the onset of CMR and the deviation of the Curie-Weiss behavior \cite{rosa2020colossal}. Our results therefore corroborate the scenario wherein the formation of trapped magnetic polarons reduces the number of free carriers, changing the environment around the Eu$^{2+}$ site concomitantly with the global properties, which in turn increases the microwave penetration. 

\begin{figure}[!ht]
\includegraphics[width=\columnwidth]{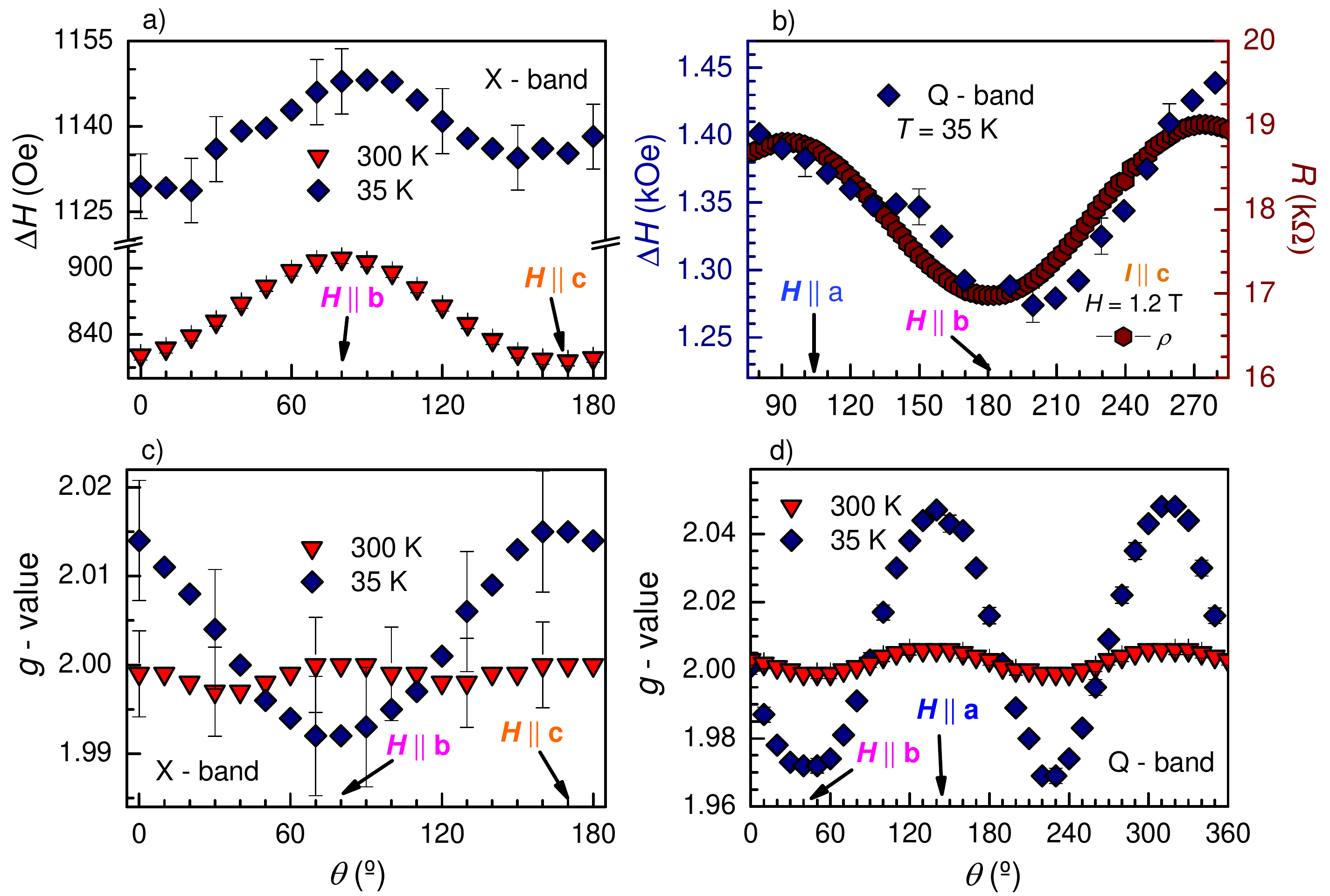}
\caption{Anisotropy of the a) X- and b) Q-bands Eu$^{2+}$ ESR $\Delta H$. The current $I$ was applied parallel to the $c$-axis in the resistivity measurement. The brown hexagonal symbols show the resistivity anisotropy at $T$ = 35 K. c) X- and d) Q-bands Eu$^{2+}$ $g$-value anisotropy at $T$ = 300 K and 35 K. The anisotropy in X-band went from $H$ applied parallel from the $b$ to the $c$ axis, while $H$ was applied from the $b$ to the $a$ axis in Q-band.}
\label{Fig3}
\end{figure}

To uncover the anisotropy of Eu$_{5}$In$_{2}$Sb$_{6}$, we turn to the angle dependence of the Eu$^{2+}$ ESR $\Delta H$ and $g$-factor at $T$ = 35 K and 300 K, shown in Figure \ref{Fig3}. The anisotropy of $\Delta H$ in the $ab$ plane is less pronounced than its out-of-plane anisotropy, which suggests the presence of an orthorhombic contribution due to weak crystal field effects \cite{abragam2012electron,rosa2014EuPtIn4}. Nonetheless, the Eu$^{2+}$ ESR $\Delta H$ is also connected with $T_{2}$, and the resistivity anisotropy in the $ab$ plane resembles the anisotropy in $\Delta H$, as shown in fig. \ref{Fig3} b). A similar angular dependence for both physical quantities indicates the presence of anisotropic magnetic scattering stemming from anisotropic magnetic interactions between Eu$^{2+}$ ions. 
In fact, a $g$-factor anisotropy is observed even at room temperature, as shown in Figs. \ref{Fig3} c) and d).
The experimental Eu$^{2+}$ $g$-value, $g$ = $h$$\nu$/$\mu_{B}$$H_{r}$, is obtained from the analysis of the Eu$^{2+}$ ESR spectra using eq. \ref{Eq1}. 
Here $h$ is the Planck constant and $\mu_{B}$ the Bohr magneton. Eu$^{2+}$ is a localized $S$ ion ($L$ = 0), and its second-order crystal field effects cannot count for the $g$-value anisotropy \cite{barnes1981theory,abragam2012electron,urbano2004magnetic,urbano2006esr}. Therefore, such anisotropy is an indication that even at room temperature short-range magnetic interactions 
are substantial in Eu$_{5}$In$_{2}$Sb$_{6}$, a rare occurrence in localized 4$f$-electron materials. At $T = 35$~K, the anisotropy is even more pronounced. Such $g$-value anisotropy typically stems from internal fields and supports the presence of complex magnetic interactions in the paramagnetic phase of Eu$_{5}$In$_{2}$Sb$_{6}$. Demagnetization effects do not fully describe the $g$-factor anisotropy (Appendix A).

\begin{figure*}[!ht]
\includegraphics[width=2\columnwidth]{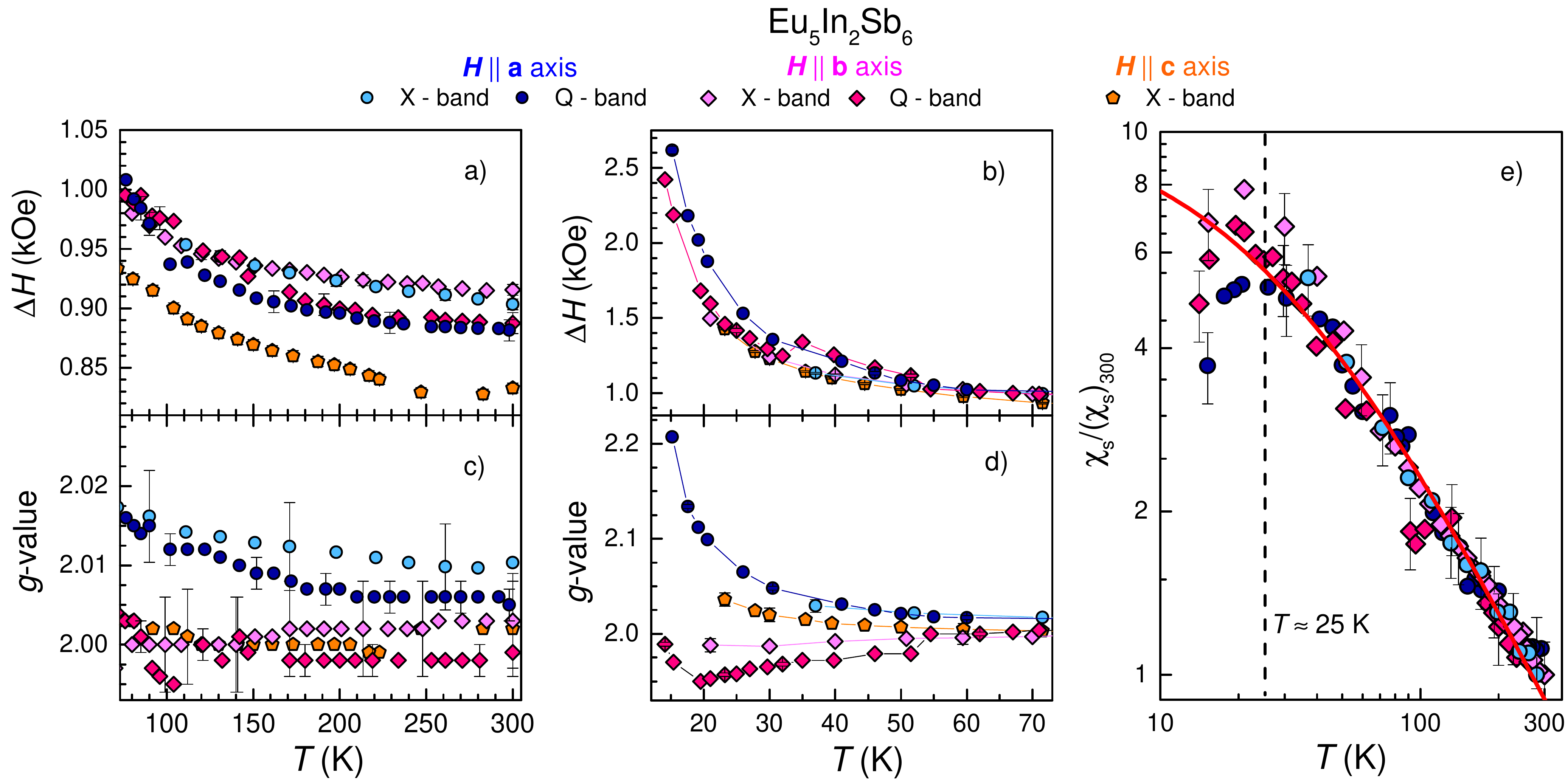}
\caption{High (70 K $\leq$ T $\leq$ 300 K) and low (15 K $\leq$ T $\leq$ 70 K) $T$-dependencies of the Eu$^{2+}$ ESR a), b) $\Delta H$ and c), d) $g$-value for X- and Q-band for different crystallographic directions. The solid lines are guides to the eye. e) Eu$^{2+}$ spin susceptibility $\chi_{s}$ normalized by its value at $T$ = 300 K for different crystallographic directions. The red solid line is a Curie-Weiss-like curve with $\theta$ = - 28 K.}
\label{Fig4}
\end{figure*}

Figures \ref{Fig4} a) and b) show the $T$-dependence of the Eu$^{2+}$ ESR $\Delta H$ for two different microwave frequencies at high and low temperatures, respectively. 
Fig. \ref{Fig4} a) shows a systematic reduction of the Eu$^{2+}$ ESR $\Delta H$ at higher $H$s (Q-band) at high temperatures. Such reduction appears even at room temperature, which can be explained by a reduction of the spin-flip scattering. According to Ref. \cite{rosa2020colossal}, the dominant coupling between 4$f$ local moments and conduction electrons at high temperatures is ferromagnetic, and the resulting ferromagnetic clusters therefore increase in size as a function of $H$ \cite{abragam2012electron,poole1971relaxation}. Although the high-temperature data show a reduction of the Eu$^{2+}$ $\Delta H$, a Korringa-like behavior of the Eu$^{2+}$ $\Delta H$ $T$-dependence is absent. A Korringa behavior is reflected in a positive linear-in-$T$ dependence of Eu$^{2+}$ $\Delta H$. This result indicates that, even at high temperatures, the internal fields due to Eu$^{2+}$ spin-spin interaction are dominant.

Notably, below $T^{\ast}$ $\sim$ 50 K the Eu$^{2+}$ ESR linewidth increases rapidly with decreasing temperature [Figure \ref{Fig4} b)]. This crossover region likely stems from the strong short-range interactions between polarons as previously suggested by the breakdown in activated behavior in resistivity, the strong AFM correlations in magnetic susceptibility, and the Schottky anomaly of the specific heat \cite{rosa2020colossal}. Because the Eu$_{5}$In$_{2}$Sb$_{6}$ structure hosts three different Eu$^{2+}$ sites, complex local magnetic fields and anisotropic Eu$^{2+}$ spin-spin interactions are expected. Anisotropic short-range magnetic interactions create different magnetic environments, which results in inhomogeneous broadening \cite{poole1971relaxation,abragam2012electron,rivadulla1999electron}. This type of broadening is expected when strong local internal magnetic fields start to play a role, which in the case of Eu$_{5}$In$_{2}$Sb$_{6}$ occurs near a magnetic transition with enhanced interactions between polarons. As a matter of comparison, at $T^{\ast}$ $\sim$ 50 K the Eu$^{2+}$ ESR line shape starts to distort for X-band, which explains the small upturn of $\alpha$ observed in this region in Fig. \ref{Fig2} b). This is another indication that inhomogeneous broadening is the responsible for such crossover region.

Figures \ref{Fig4} c) and d) show the Eu$^{2+}$ ESR $g$-value as a function of temperature. At high temperatures, there is a systematic reduction of the Eu$^{2+}$ $g$-value as the frequency is increased from X- to Q-band. This reduction stems from short-range interactions between Eu$^{2+}$ ions, which generates local AFM fields. Positive $g$-values are related to FM internal fields, whereas negative $g$-values result from AFM interactions. Therefore, a systematic reduction of the $g$-value as a function of $H$ will result in an increase of AFM internal fields as a function of the microwave frequency. This result supports the magnetic polaron scenario and shows microscopically that the interaction between polarons is AFM.

Because the $g$-value is an important probe of local magnetic interactions, a closer look at its temperature dependence in the paramagnetic phase can provide insights into the magnetic structure of Eu$_{5}$In$_{2}$Sb$_{6}$ below $T_{N1}$. Fig.~\ref{Fig4}~d) shows the low-temperature behavior of the Eu$^{2+}$ $g$-value. Along the 
$c$ and $a$ axes, the $g$-value increases on cooling, whereas a decrease is observed along the $b$ axis. The $g$-value temperature dependence for $T$ $\geq 20 K$ sheds light on the complex magnetic structure of Eu$_{5}$In$_{2}$Sb$_{6}$ that arises from competing AFM and FM interactions in the $ab$ plane where the spins lie. 
In particular, a FM component along $a$ is consistent with the small net moment and the increase of the magnetic susceptibility on cooling through $T_{N1}$ for fields applied along the $a$ axis \cite{rosa2020colossal}. This magnetic structure differs from the proposed axion insulator scenario \cite{rosa2020colossal}, which invites spectroscopic measurements to determine
the magnetic structure of Eu$_{5}$In$_{2}$Sb$_{6}$ as well as further calculations taking into account the experimentally-determined magnetic structure.

An alternative scenario to the Eu$^{2+}$ spin dynamics could rely on the opening of a spin gap \cite{little2017antiferromagnetic,ponomaryov2017unconventional,ponomaryov2020nature,miksch2021gapped} rather than magnetic polarons physics. In general, such opening is less expected in Eu$^{2+}$-based compounds due to their negligible spin-orbit coupling. The experimental signature of a spin gap should be observed in the Eu$^{2+}$ spin susceptibility $\chi_{s}$ \cite{miksch2021gapped,chabre2005low,zorko2004x,prokofiev1998magnetic}. Fig.~\ref{Fig4}~e) show $\chi_{s}$ $\propto$ $g^{2}I_{ESR}$, where $I_{ESR}$ is the Eu$^{2+}$ ESR intensity, as a function of temperature - $\chi_{s}$ is normalized by its value at 300 K. $\chi_{s}$ shows a Curie-Weiss-like behavior ($\chi_{s}$ $\propto$ $1/(T - \theta)$), where $\theta$ is the Curie-Weiss temperature - red solid line with $\theta$ = - 28 K. The obtained $\theta$ is at the same magnitude of the $\theta$ obtained by magnetization ($\sim$ 30 K) \cite{rosa2020colossal}, but with opposite sign. The negative sign reflects the presence of an AFM ground state. For $T$ $\leq$ 25 K there is a clear reduction of $\chi_{s}$. In addition, the opening of a spin gap, if present, would also be evidenced in the ESR spin dynamics. The Eu$^{2+}$ $\Delta H$ and $g$-value greatly increase at this temperature range for all directions [Figs.~\ref{Fig4}~b), d)]. More specifically, there is an increase of the $g$-value for $H$ parallel to the $b$ axis for $T$ $\leq$ 20 K, which could be interpreted as another signature of a spin gap opening; however, other mechanisms such as short-range interactions or strong magnetic fluctuations could also be associated to all these signatures. Nonetheless, such signatures for $T$ $\leq$ 20 K clearly show that the effects at higher temperatures are not associated with the opening of a spin gap or strong magnetic fluctuations, but most likely originates from the magnetic polarons physics.

Our results shed microscopic light on three basic energy scales: $T'$ $\sim$ 200 K, $T^{\ast}$ $\sim$ 50 K, and $H$-dependence. At $T'$ $\sim$ 200 K, trapped magnetic polarons start to become sizable, which cause a decrease in carrier density. An increase in $\delta$ is therefore observed on cooling and, consequently, a change of the Eu$^{2+}$ line shape takes place. Conversely, the effect of $H$ is to polarize more carriers around the Eu$^{2+}$ magnetic ions, which gives rise to a spin-dependent scattering. As a result we obtain a reduction of the spin-flip scattering, which causes unusual narrowing of the Eu$^{2+}$ line width, and, macroscopically, a CMR \cite{rosa2020colossal}. Though decreasing the temperature at a constant $H$ also increases the size of the magnetic polarons, this effect does not dramatically change the spin-dependent scattering, as evidenced by the Eu$^{2+}$ ESR $\Delta H$ $T$-dependence for $T$ $\geq$ 150 K. If the spin-flip scattering were to decrease significantly, we would obtain an increase of the difference of the Eu$^{2+}$ X- and Q-band linewidths as a function of temperature, which is not the case [Fig. \ref{Fig4} a)]. At $T^{\ast}$ $\sim$ 50 K, the trapped polarons start to interact, which results in different magnetic states at the Eu$^{2+}$ site and the appearance of an inhomogeneous broadening. Note that magnetic polarons are still localized at $T^{\ast}$ because an insulating Lorentzian Eu$^{2+}$ line shape is still observed and the resistivity still increases on cooling. This is consistent, for instance, with the appearance of AFM correlations in the magnetic susceptibility at $T^{\ast}$ $\sim$ 40 K \cite{rosa2020colossal}. Finally, at $T_{N1}$ the polarons become delocalized, which results in a significant drop in the resistivity of Eu$_{5}$In$_{2}$Sb$_{6}$ \cite{rosa2020colossal}.

\section{\label{sec:conclusion}IV. Conclusion}

In summary, we performed electron spin resonance in the antiferromagnetic insulator Eu$_{5}$In$_{2}$Sb$_{6}$ using different microwave frequencies in the temperature range of 15 K $\leq$ $T$ $\leq$ 300 K. Our Eu$^{2+}$ ESR spin dynamics analysis reveals a decrease of the Eu$^{2+}$ ESR linewidth when going from low  to higher $H$s, consistent with the presence of ferromagnetic clusters. Below $T'$ $\sim$ 200 K, the change of the asymmetry of the Eu$^{2+}$ line shape reveals a marked change of $\delta$ due to the increase of the magnetic polaron size. At $T^{\ast}$ $\sim$ 50 K, strong inter-polaron antiferromagnetic interactions cause an inhomogeneous broadening and $g$-value changes due to anisotropic short-range interactions. Our microscopic analysis sheds light onto the complex magnetic structure of Eu$_{5}$In$_{2}$Sb$_{6}$ and invites other spectroscopic measurements to investigate its putative axion insulating phase.

\begin{acknowledgments}

We thank J. Santos Rego, M. C. Rahn, M. V. Ale Crivillero, S. Wirth, and R. Urbano for fruitful discussions. This work was supported by FAPESP\ (SP-Brazil) Grants No 2020/12283-0, 2018/11364-7, 2017/10581-1, National Council of Scientific and Technological Development - CNPq Grants No 442230/2014-1 and 304649/2013-9, CAPES, FINEP-Brazil and Brazilian Ministry of Science, Technology and Innovation. Work at Los Alamos National Laboratory (LANL) was performed under the auspices of the U.S. Department of Energy, Office of Basic Energy Sciences, Division of Materials Science and Engineering. Part of this work was supported by the Center for Integrated Nanotechnologies, an Office of Science User Facility operated for the U.S. Department of Energy Office of Science.

\end{acknowledgments}

\section{\label{sec:AppI}Appendix A: Demagnetization effects}

Demagnetization effects are important whenever systems are uniformly magnetized \cite{kittel1948theory}. In other words, if the sample size is small compared to the skin depth of the microwave, we should look for such effects. The anisotropy of the $g$-value could be a result of the shape of the crystal, and not necessarily due to an intrinsic magnetic anisotropy \cite{urbano2006esr,wellm2020magnetic}.

In our case, the skin depth at $T$ = 300 K is $\delta^{300 K}$ = 0.2 mm, while for $T$ = 35 K it is $\delta^{35 K}$ = 25 mm. Due to our sample dimensions (0.5 mm x 0.5 mm x 3 mm), we are only going to evaluate the effects at $T$ = 35 K. The resonance frequency in the presence of demagnetization effects is given by

\begin{equation}
\omega _{0}^{2} = \gamma^{2} \left [ H + \left ( N_{y} - N_{z} \right ) M  \right ]\left [ H + \left ( N_{x} - N_{z} \right ) M \right ],
\label{Eqs1}
\end{equation}
where $\gamma$ = $e/mc$ is the gyromagnetic factor, $H$ the applied magnetic field, $M$ the magnetization and $N_{i}$ ($i$ = $x, y, z$) the demagnetizing factors, with $\sum_{i} N_{i} = 1$ \cite{kittel1948theory,wellm2020magnetic}. The simplest case is the thin plate, where $N_{x}$ = $N_{y}$ = 0 and $N_{z}$ = 1 when the magnetic field is perpendicular to the plate \cite{kittel1948theory}.

For a paramagnetic material we have $M = \chi H$, where $\chi$ is the magnetic susceptibility. We also note that

\begin{equation}
g = \frac{\hbar \omega_{0}}{\mu_{B} H} = \frac{g_{e}}{H} \sqrt{\left [ H + \left ( N_{y} - N_{z} \right ) M  \right ]\left [ H + \left ( N_{x} - N_{z} \right ) M \right ]},
\label{Eqs2}
\end{equation}
where $\mu_{B}$ is the Bohr magneton, $\hbar$ the Planck constant divided by 2$\pi$ and $g_{e}$ $\approx$ 2.002 the free electron $g$-value. Using eq. \ref{Eqs2} into \ref{Eqs1} and the fact that we are probing the paramagnetic phase, we obtain

\begin{equation}
g = g_{e} \sqrt{\left [ 1 + \left ( N_{y} - N_{z} \right ) \chi  \right ]\left [ 1 + \left ( N_{x} - N_{z} \right ) \chi \right ]}.
\label{Eqs3}
\end{equation}

In order to analyze the $N_{i}$'s, as a first approximation, we assumed a rectangular prism form of the measured crystals. A complete description of the method to obtain the demagnetizing factors for a rectangular prism is described in ref. \cite{fukushima1998volume}. We obtain $N_{x}$ = $N_{z}$ $\approx$ 0.46, $N_{y}$ $\approx$ 0.08 when $H$ is parallel to the $a$ axis, $N_{y}$ = $N_{z}$ $\approx$ 0.46, $N_{x}$ $\approx$ 0.08 for $H$ parallel to the $b$ axis and, finally, $N_{x}$ = $N_{y}$ $\approx$ 0.46, $N_{z}$ $\approx$ 0.08 when $H$ is parallel to the $c$ axis.

Using the value of the molar magnetic susceptibility $\chi_{m}$ reported in ref. \cite{rosa2020colossal}, we can calculate $\chi$ = $\chi_{m}$ $D$/$M_{m}$, where $D$ is the density and $M_{m}$ the molar mass. Using $D$ = 6.77 g/cm$^{3}$ and $M_{m}$ = 1719.94 g/mol, we obtain $\chi$ = 0.011. Finally, using eq. \ref{Eqs3}, the demagnetizing contribution to the $g$-value anisotropy is $\Delta g^{ab}$ = $g^{H \| a}$ - $g^{H \| b}$ = 0 and $\Delta g^{cb}$ = $g^{H \| c}$ - $g^{H \| b}$ = 0.013.

Experimentally we obtained $\Delta g_{exp}^{ab}$ = 0.075(5) and $\Delta g_{exp}^{cb}$ = 0.023(5). Although imperfections of the crystal could result in a finite $\Delta g^{ab}$, the experimental value is substantial and above any experimental uncertainty, especially when comparing with $\Delta g^{cb}$, for example. This result clearly shows that the $g$-value anisotropy is related to an intrinsic magnetic anisotropy, and not due to demagnetizing effects. Moreover, $\Delta g_{exp}^{cb}$ is almost two times larger than the demagnetizing factor. Although the anisotropy out of the plane has an influence of these effects, intrinsic magnetic anisotropy is also present.


%

\end{document}